\documentclass[12pt]{article}%
\pdfoutput=1
\usepackage[nosort]{cite}
\usepackage{graphicx}
\usepackage{multicol}
\usepackage{amsfonts}
\usepackage{amssymb}
\usepackage{amsmath}
\usepackage{singleauthor}
\usepackage{afterpage}
\usepackage{setspace}
\usepackage{verbatim}
\usepackage{slashed}
\usepackage{color}
\usepackage{longtable}
\usepackage{float}
\usepackage{subcaption}
\usepackage{epsfig}
\usepackage{bm}
\usepackage{enumerate}
\usepackage{epstopdf}
\usepackage[enableskew, vcentermath]{youngtab}
\usepackage{adjustbox}
\newsavebox{\mysavebox}

\usepackage[margin=1in]{geometry}
\usepackage{titletoc}%
\usepackage{hyperref}

\newcommand{\ba}{\begin{eqnarray}}
\newcommand{\ea}{\end{eqnarray}}

\newcommand{\e}{\mathrm{e}}
\newcommand{\rmd}{\mathrm{d}}
\newcommand{\df}{\mathrel{:=}}

\def\be{\begin{equation}}
\def\ee{\end{equation}}

\makeatletter \@addtoreset{equation}{section} \makeatother

\begin{document}

\date{\today}

\title{Dimensional Reduction \\and (Anti) de Sitter Bounds}

\institution{BERKELEY}{Physics Department, University of California, Berkeley CA 94720 USA}

\authors{Tom Rudelius\worksat{\BERKELEY}\footnote{e-mail: {\tt rudelius@berkeley.edu}}}

\abstract{
Dimensional reduction has proven to be a surprisingly powerful tool for delineating the boundary between the string landscape and the swampland. Bounds from the Weak Gravity Conjecture and the Repulsive Force Conjecture, for instance, are exactly preserved under dimensional reduction. Motivated by its success in these cases, we apply a similar dimensional reduction analysis to bounds on the gradient of the scalar field potential $V$ and the mass scale $m$ of a tower of light particles in terms of the cosmological constant $\Lambda$, which ideally may pin down ambiguous $O(1)$ constants appearing in the de Sitter Conjecture and the (Anti) de Sitter Distance Conjecture, respectively. We find that this analysis distinguishes the bounds $|\nabla V|/V \geq \sqrt{4/(d-2)}$, $m \lesssim |\Lambda|^{1/2}$, and $m \lesssim |\Lambda|^{1/d}$ in $d$-dimensional Planck units. The first of these bounds precludes accelerated expansion of the universe in Einstein-dilaton gravity and is almost certainly violated in our universe, though it may apply in asymptotic limits of scalar field space. The second bound cannot be satisfied in our universe, though it is saturated in supersymmetric AdS vacua with well-understood uplifts to 10d/11d supergravity. The third bound likely has a limited range of validity in quantum gravity as well, so it may or may not apply to our universe. However, if it does apply, it suggests a possible relation between the cosmological constant and the neutrino mass, which (by the see-saw mechanism) may further provide a relation between the cosmological constant problem and the hierarchy problem. We also work out the conditions for eternal inflation in general spacetime dimensions, and we comment on the behavior of these conditions under dimensional reduction.
}

\maketitle

\tableofcontents

\enlargethispage{\baselineskip}

\setcounter{tocdepth}{2}

\section{Introduction}\label{sec:INTRO}

The search for universal features of quantum gravity--also known as the swampland program \cite{Vafa:2005ui,Ooguri:2006in}--has seen a resurgence in recent years. Strong evidence has been given in favor of certain conjectured properties of quantum gravities (so-called ``swampland conjectures''), some longstanding conjectures have been discarded as counterexamples have emerged, and many seemingly-unrelated aspects of physics and mathematics have been connected through an ever-growing swampland web.

But the swampland program also faces some serious difficulties. First and foremost is the infamous swampland tradeoff: conjectures that tend to have more evidence in their favor tend to have less interesting applications to cosmology and phenomenology, whereas conjectures that have significant implications for observable physics tend to be more speculative. Related to this is a lack of precision: many swampland statements are plagued by squiggles ($\sim$, $\lesssim$, $\gtrsim$) and $O(1)$ constants to be determined later. These squiggles make the conjectures difficult to test, and they suggest that we may not yet understand the underlying physics responsible for such conjectures. Without this, it is difficult to say when exactly the conjectures may be expected to hold, and when they may cease to be valid.

The goal of this paper is to sharpen some of these more speculative swampland conjectures, primarily through a mechanism that has successfully sharpened swampland conjectures in the past: dimensional reduction. As we will see in Section \ref{sec:WGC} below, $O(1)$ factors in the Weak Gravity Conjecture (WGC) bound and the Repulsive Force Conjecture (RFC) bound can be fixed precisely by demanding that these bounds should be exactly preserved under dimensional reduction from $D$ to $d=D-1$ dimensions: the bound will be saturated in $d$ dimensions if and only if it is saturated in $D$ dimensions. In these cases, dimensional reduction succeeds in selecting bounds that are both physically meaningful and seemingly-universal: all known quantum gravities satisfy both the WGC and the RFC, and compelling evidence has been provided that these conjectures should hold more generally.

Of course, this does not prove that \emph{any} bound that is preserved under dimensional reduction is necessarily a universal constraint on quantum gravity. But the success of dimensional reduction in the above cases suggests that it may serve as a useful diagnostic for distinguishing certain $O(1)$ values in conjectured bounds when previous arguments have left us with ambiguity.

Relatedly, this paper will not address the ultimate question of whether certain proposed conjectures are universally true in quantum gravity. It seems very likely that many of these conjectures are true within a limited domain of validity. For instance, the relationship between the masses of a tower of light particles and the cosmological constant proposed in the AdS Distance Conjecture \cite{Lust:2019zwm} is likely true in infinite families of quantum gravities, each member of which is distinguished by some discrete flux parameter. The lower bound on the gradient of a scalar field potential proposed in the de Sitter Conjecture \cite{Obied:2018sgi} is likely true in asymptotic regions of scalar field space, and relatedly eternal inflation is likely forbidden in such regions \cite{Rudelius:2019cfh}. The bounds we will obtain in this paper do not necessarily apply to all quantum gravities--in fact, some of them are violated even in our own universe--but they may point us to interesting, universal behavior in certain limits of quantum gravity. At the least, they give us a mark to aim for and to explore further. 

The main results and structure of our paper is as follows: in Section \ref{sec:WGC}, we will review how dimensional reduction distinguishes the WGC and RFC bounds, fixing their $O(1)$ factors. 

In Section \ref{sec:LIGHT}, we will apply a similar dimensional reduction analysis to bounds on the masses of light particles in terms of the cosmological constant of the form
\be
m \lesssim |\Lambda_d|^{\alpha_d} M_d^{1- d \alpha_d}\, ,
\ee
where $\Lambda_d$ and $M_d$ are the cosmological constant and the Planck mass in $d$ dimensions, respectively. In \cite{Lust:2019zwm}, it was conjectured that a tower of light particles should satisfy this bound in any AdS (and perhaps dS) vacuum. We will see that dimensional reduction picks out $\alpha_d = 1/2$ and $\alpha_d = 1/d$ as special values of $\alpha_d$, depending on whether or not we include the contribution from 1-loop Casimir effects. The former value is saturated in known supersymmetric compactifications \cite{Lust:2019zwm} and was previously singled out in \cite{Andriot:2020lea}, whereas the latter is interesting from the perspective of the standard model, since neutrinos have a mass of roughly $m_\nu \sim \Lambda_4^{1/4}$. If this mass comes from the see-saw mechanism, $m_\nu \sim y^2 v^2 / \Lambda_{\text{UV}}$, where $y$ is the Yukawa coupling, $v$ is the Higgs vev, and $\Lambda_{\text{UV}}$ is some UV scale, this bound further becomes
\be
v^2  \lesssim \Lambda_4^{1/4}  \Lambda_{\text{UV}} / y\,,
\label{vevbound}
\ee
which bounds the Higgs vev in terms of the cosmological constant and therefore relates the hierarchy problem to the cosmological constant problem. In particular, if $\Lambda_{\text{UV}} \sim M_{\text{GUT}}$, $y \sim 0.1$, then \eqref{vevbound} becomes $v \lesssim 100$ GeV, which is not far from the measured value $v = 246$ GeV. (Note that we have swept a number of $O(1)$ factors here under the rug.)

In Section \ref{sec:ETERNAL}, we will derive constraints for eternal inflation in $d \geq 4$ spacetime dimensions by solving the associated Fokker-Planck equation analytically, generalizing the $d=4$ result of \cite{Rudelius:2019cfh}. We will find that eternal stochastic inflation occurs only if the following conditions are satisfied on the first and second derivatives of the potential:
\begin{align}
\frac{|\nabla V|^2}{V^{(d+2)/ 2} } &<  \frac{2 (d-1)^3}{\pi \Omega_{d-2} } \left(  \frac{ 2  }{ (d-1)(d-2) M_d^{d-2} }   \right)^{(d+2)/2}\,, \label{eiintro} \\
\frac{V''}{V} &> -\frac{2 (d-1)}{d-2} \frac{1}{M_d^{d-2}} \,,
\end{align}
where $\Omega_{d-2} = {2 \pi^{(d-1)/2}/\Gamma(\frac{d-1}{2})}$. In multifield landscapes, $V''$ above should be replaced by a sum over the negative eigenvalues of the Hessian of the potential.

In Section \ref{sec:DS}, we will consider bounds of the form 
\be
\frac{ |\nabla V_d|  }{ V_d^{\gamma_d}  }  \geq c_d M_d^{d(1-\gamma_d)-(d-2)/2}\,.
\label{introeq}
\ee
Preservation under dimensional reduction distinguishes $\gamma_d = 1$ as a special value of $\gamma_d$, in agreement with the de Sitter Conjecture (dSC) bound of \cite{Obied:2018sgi} and the Trans-Planckian Censorship Conjecture (TCC) bound. It further selects $c_d = \sqrt{4/(d-2)}$ as a special value of $c_d$, in which case \eqref{introeq} corresponds exactly to the condition required to avoid accelerated expansion of the universe. Since this particular value of $c_d$ is larger than the coefficient $c_d = 2/\sqrt{(d-1)(d-2)}$ that appears in the TCC bound, we learn that a theory satisfying the TCC bound will still satisfy the TCC bound after dimensional reduction. Likewise, since our value $\gamma_d = 1$ is smaller than the value $\gamma_d = (d+2)/4$ appearing in the eternal inflation bound \eqref{eiintro}, we see that the dimensionally reduced theory will not lead to eternal inflation in asymptotic regions of scalar field space; eternal inflation can occur only if the radion is stabilized. 

We end in Section \ref{sec:CONC} with conclusions and directions for future research.


\section{Review: Dimensional Reduction and Weak Gravity Conjectures}\label{sec:WGC}

In this section, we review how the Weak Gravity Conjecture (WGC) \cite{ArkaniHamed:2006dz} bound and the Repulsive Force Conjecture (RFC) \cite{Palti:2017elp} bounds are preserved under dimensional reduction.

Our starting point is an Einstein-Maxwell-dilaton action for a $P$-form gauge field $A_{\mu_1 \ldots \mu_P}$ in $D =d+1$ dimensions:
\be
S = \frac{1}{2\kappa_D^2} \int \rmd^D x \sqrt{-g} \left({\cal R}_D - \frac{1}{2} (\nabla \phi)^2\right) - \frac{1}{2e_{P;D}^2} \int \rmd^D x \sqrt{-g} \e^{-\alpha_{P;D} \phi} F_{P+1}^2 \,. \label{eq:generalaction}
\ee
Here, $F_{P+1} = \rmd A_P$ is the field strength for  $P$-form gauge field, and
\be
F_q^2 \df \frac{1}{q!} F_{\mu_1 \ldots \mu_q} F^{\mu_1 \ldots \mu_q} \,.
\ee
Note that the action \eqref{eq:generalaction} is rather special and does not describe the low-energy theory of a generic quantum gravity, but it will suffice for our purposes to restrict our attention to this case.

We define
\be
\frac{1}{\kappa_D^2} \df M_D^{D-2},
\ee
where $M_D$ is the reduced Planck mass in $D$ dimensions. With this convention, the WGC bound for a $(P-1)$-brane of quantized charge $q$ and tension $T_P$ is given by:
\begin{equation}
    e_{P;D}^2 q^2 M_{D}^{D-2} \geq \left[ \frac{\alpha_{P;D}^2}{2} + \frac{P(D-P-2)}{D-2} \right] T_P^2 \df \gamma^{-1}_{P; D}(\alpha) T_P^2 \,.
    \label{eq:extremalitybound}
\end{equation}
The RFC bound in this theory is given by 
\begin{equation}
    e_{P;D}^2 q^2 M_{D}^{D-2} \geq 2 (\partial_\phi T_P)^2 + \frac{P(D-P-2)}{D-2}  T_P^2 \, ,
     \label{eq:repulsion}
\end{equation}
where $\partial_\phi T_P \df \partial T_P / \partial \phi$ is the partial derivative of the brane tension with respect to the dilaton $\phi$, holding the Planck scale fixed. A $(P-1)$-brane that satisfies the WGC bound is said to be superextremal: its charge-to-mass ratio is greater than that of an extremal black brane. A $(P-1)$-brane that satisfies the RFC bound is self-repulsive: two such parallel branes separated by a parametrically large distance will feel a repulsive force between them.

We now show how these bounds are preserved under dimensional reduction. We consider a dimensional reduction ansatz of the form,
\be
\rmd s^2 = \e^{\frac{\lambda(x)}{d-2}} \rmd{\hat s}^2(x) + \e^{-\lambda(x)} \rmd y^2,
   \label{eq:dimredansatz}
\ee
where $y \sim y + 2 \pi R$. This ansatz is chosen so that the dimensionally reduced action is in Einstein frame, i.e., there is no kinetic mixing between $\lambda$ and the $d$-dimensional metric. For simplicity, we do not include a Kaluza-Klein photon, but we do include a massless radion $\lambda(x)$, which controls the radius of the circle. Under such a dimensional reduction, the Einstein-Hilbert term reduces as
\be
 \frac{1}{2\kappa_D^2} \int \rmd^D x \sqrt{-g} {\cal R}_D \rightarrow  \frac{1}{2\kappa_d^2} \int \rmd^d x \sqrt{- \hat g}  {\cal R}_d  - \frac{1}{2}  \int \rmd^d x \sqrt{- \hat g} \, G_{\lambda \lambda} (\nabla \lambda)^2 \,,
\ee
where
\begin{align}
\frac{1}{\kappa_d^2} &= M_d^{d-2} = (2 \pi R) M_D^{D-2} \,, \label{Ddconstants} \\
G^{(d)}_{\lambda \lambda} &= \frac{(d-1)}{4\kappa_d^2 (d-2)} = M_d^{d-2} \frac{d-1}{4(d-2)}\,.
\label{radionkineticterm}
\end{align}

The $P$-form in $D$ dimensions gives both a $P$-form and a $p \df P-1$-form in $d$ dimensions. The former comes from taking all of the legs of the $P$-form to lie along noncompact directions, while the latter comes from taking one of the legs of the $P$-form to wrap the compact circle. Similarly, a $(P-1)$-brane charged under the $P$-form descends to both a $(P-1)$-brane and a $(p-1)$-brane, charged under the respective forms. For simplicity, we present only the former here: the latter computation is analogous.

Consider then the $P$-form gauge field in $d$ dimensions. The associated gauge coupling is given by
\begin{equation}
    e_{P;d}^2 = \frac{1}{2 \pi R}e_{P;D}^2 \e^{\frac{P \lambda}{d-2}} \,.
\end{equation}
The tension of a $(P-1)$-brane transverse to the compact circle is given by
\begin{equation}
T_P^{(d)} = \e^{\frac{P \lambda}{2(d-2)}} T_P^{(D)}\,.
\label{eq:TPd}
\end{equation}
Upon reduction, the radion $\lambda$ and dilaton $\phi$ each couple exponentially to the Maxwell term in the action, as in \eqref{eq:generalaction}. We may then redefine the dilaton as the linear combination of $\lambda$ and $\phi$ that couples to the Maxwell term. This effectively shifts the coupling $\alpha$ appearing in the WGC bound to
\begin{equation}
    \alpha_{P;d}^2 = \alpha_{P;D}^2 + \frac{2 P^2}{(d-1)(d-2)}.
\end{equation}
Plugging this into the WGC bound (\ref{eq:extremalitybound}) with $D \rightarrow d$, we find
\begin{align}
    e_{P;d}^2 q^2 M_{d}^{d-2} \geq \left[ \frac{\alpha_{P;d}^2}{2} + \frac{P(d-P-2)}{d-2} \right] T_{P}^2
    = \left[ \frac{\alpha_{P;D}^2}{2} + \frac{P(D-P-2)}{D-2} \right] T_{P}^2.
\end{align}
The right-hand side of this bound matches that of \eqref{eq:extremalitybound}. This shows that the WGC bound has been exactly preserved by the dimensional reduction process. An analogous result holds for the case of decreasing $P \rightarrow P-1$ and for Kaluza-Klein modes when $P=1$ provided we include a Kaluza-Klein photon in the dimensional reduction ansatz of \eqref{eq:dimredansatz} \cite{Heidenreich:2015nta}: in all such cases, the WGC bound is exactly preserved under reduction.

The RFC bound in $d$ dimensions is then given by 
\begin{equation}
e_{P;d}^2 q^2 \geq  \frac{1}{2 \pi R} (\partial_\phi T_P^{(d)})^2 + (G^{(d)})^{\lambda\lambda} (\partial_\lambda T_P^{(d)})^2 - \frac{P(d-P-2)}{d-2}\frac{\big(T_P^{(d)}\big)^2}{M_d^{d-2}}.  \label{eq:dimreducedselfforce}
\end{equation}
Using \eqref{radionkineticterm} and \eqref{eq:TPd}, we have
\begin{equation}
(G^{(d)})^{\lambda\lambda} (\partial_\lambda T_P^{(d)})^2 = \frac{4(d-2)}{M_d^{d-2}(d-1)} \left(\frac{P}{2(d-2)}\right)^2 \big(T_P^{(d)}\big)^2 = \frac{P^2}{(d-1)(d-2)} \frac{\big(T_P^{(d)}\big)^2}{M_d^{d-2}}\,.
\end{equation}
This combines with the last term in (\ref{eq:dimreducedselfforce}) to give
\begin{equation}
\left[\frac{P^2}{(d-1)(d-2)}  + \frac{P(d-P-2)}{d-2}\right] \frac{\big(T_P^{(d)}\big)^2}{M_d^{d-2}} = \frac{P(D-P-2)}{D-2} \frac{\big(T_P^{(d)}\big)^2}{M_d^{d-2}}\,.
\end{equation}
With this, we see that the RFC bound in $d$ dimensions \eqref{eq:dimreducedselfforce} is exactly the same as the RFC bound in $D$ dimensions \eqref{eq:repulsion}. This result may be generalized to multiple scalar fields, multiple gauge fields, and general scalar field couplings. It holds also for the case of $P$ reduced to $p = P-1$ and for Kaluza-Klein modes when $P=1$ \cite{Heidenreich:2019zkl}: in all cases, the RFC bound is exactly preserved under dimensional reduction.

Within the string landscape, there are no known counterexamples to either the WGC or the RFC. Many examples have been confirmed to satisfy these conjectures \cite{ArkaniHamed:2006dz, Heidenreich:2019zkl, Heidenreich:2016aqi, Lee:2018urn,Klaewer:2020lfg,Lee:2019tst,Lee:2018spm, Grimm:2019wtx, Gendler:2020dfp, Cota:2020zse}, and a number of arguments suggest that they should hold more generally \cite{Cheung:2014ega, Hod:2017uqc, Heidenreich:2017sim, Cheung:2018cwt, Urbano:2018kax, Montero:2018fns}.

The takeaway lesson here is that preservation under dimensional reduction can be a useful tool for identifying universal, physically meaningful behavior of quantum gravities. The value of $\gamma_{P; D}$ in \eqref{eq:extremalitybound} is distinguished in that it dictates the extremality bound for charged black branes in the theory of \eqref{eq:generalaction}, and it is also distinguished in that it is preserved under dimensional reduction. Similarly, the bound \eqref{eq:repulsion} is distinguished by the fact that it dictates the self-repulsiveness of a brane at long distances, and simultaneously it is distinguished in that it is preserved under dimensional reduction. Both of these bounds are further distinguished by the fact that they seem to be satisfied in all quantum gravity theories. We see that, at least in these two cases, dimensional reduction picks out universal, physically meaningful constraints on quantum gravities.


\section{Light Particles and the Cosmological Constant}\label{sec:LIGHT}

Suppose that a family of AdS or dS quantum gravities contain a tower of light particles whose masses satisfy
\begin{equation}
m_n \lesssim n |\Lambda_D|^{\alpha_D} M_D^{1 - D \alpha_D} \,, ~~~~~ n \in \mathbb{Z}_{> 0}.
\label{massvcc}
\end{equation}
Such towers of light particles occur, for instance, in all known AdS vacua of string theory, typically arising as Kaluza-Klein modes of some compactified dimensions. Such towers were discussed in detail in \cite{Lust:2019zwm}, where it was further conjectured that there exists a universal, $O(1)$ value of $\alpha_D$ that is obeyed in all (A)dS vacua of string theory. This conjecture was called the (A)dS Distance Conjecture ((A)dSDC).

We now apply a dimensional reduction analysis to the bound \eqref{massvcc}. Our starting point is the $D$-dimensional action:
\be
S =  \int \rmd^D x \sqrt{-g}  \left[ \frac{M_D^{D-2}}{2}{\cal R}_D - \Lambda_D  -  \frac{1}{2} \sum_n  \left(  (\nabla \phi_n)^2  + m_n^2 \phi_n^2 \right) \right] \,,
 \label{eq:AdSaction}
\ee
where $\phi_n$ represents the $n$th particle in the tower. We have written the action as if these particles are scalar fields, but this assumption is not necessary.

We now perform a dimensional reduction to $d =D-1$ dimensions using the same ansatz as above:
\be
\rmd s^2 = \e^{\frac{\lambda(x)}{d-2}} \rmd{\hat s}^2(x) + \e^{-\lambda(x)} \rmd y^2,
   \label{eq:dimredansatz2}
\ee
The resulting action in $d$ dimensions takes effectively the same form as \eqref{eq:AdSaction}:
\be
S = \int \rmd^d x \sqrt{- \hat g}  \left[ \frac{M_d^{d-2}}{2 } \left( {\cal R}_d  -  \frac{d-1}{4(d-2)} (\nabla \lambda)^2 \right) - \Lambda_d -  \frac{1}{2} \sum_n  \left(  (\nabla \phi_n)^2  + (m_n^{(d)})^2 \phi_n^2 \right)  \right] \,,
 \label{eq:AdSactiond}
\ee
At a classical level, the parameters are related by
\be
M_d^{d-2} = (2 \pi R) M_D^{D-2} \,, ~~~~\Lambda_d = (2 \pi R) \Lambda_D \e^{\frac{\lambda}{d-2}}\,,~~~~ m_n^{(d)} = m_n^{(D)} \e^{\frac{\lambda}{2(d-2)}}\,.
\label{paramred}
\ee
There will also be Kaluza-Klein modes, which we have ignored. For simplicity, we assume here that $\langle \lambda \rangle = 0$: we can ensure this by shifting $\lambda \rightarrow \lambda - \langle \lambda \rangle$, if necessary.

From this, we see that if the masses $m_n^{(D)}$ satisfy \eqref{massvcc} in $D$ dimensions, then the masses $m_n^{(d)}$ will satisfy
\begin{equation}
m_n \lesssim n |\Lambda_d|^{\alpha_d} M_d^{1 - 2 \alpha_d} \,, ~~~~~ n \in \mathbb{Z}_{> 0}\,,
\label{massvccd}
\end{equation}
provided
\begin{equation}
|\Lambda_d|^{\alpha_d} M_d^{1 - d \alpha_d} \geq |\Lambda_D|^{\alpha_D} M_D^{1 - D \alpha_D}  = |\Lambda_d|^{\alpha_D} M_d^{\frac{d-2}{d-1} (1 - D \alpha_D) }  (2 \pi R)^{\frac{1 }{ D-2 }( 2 \alpha_D - 1)} \,.
\label{LambdadD}
\end{equation}
From this, we see that the value $\alpha_d = 1/2$ is distinguished by dimensional reduction: for this particular value, the $R$ dependence cancels, and the bound \eqref{massvcc} is exactly preserved under dimensional reduction. As pointed out in \cite{Lust:2019zwm}, the value $\alpha_d = 1/2$ is also distinguished physically in that it is saturated in all known string/M-theory backgrounds that can be understood as 10 or 11-dimensional solutions. The authors further conjecture that the bound \eqref{massvcc} should be satisfied with $\alpha_d = 1/2$ for all supersymmetric AdS vacua. This value of $\alpha_d$ was also singled out it \cite{Andriot:2020lea}, where it was connected to the dSC, the TCC, and the Swampland Distance Conjecture (SDC) \cite{Ooguri:2006in}. Once again, dimensional reduction has distinguished a bound of physical interest.

It is worth pointing out that this bound cannot be satisfied in our own universe, however. A tower of particles of this mass would run in loops, correct the graviton propagator, and lead gravity to become strongly coupled at the ``species bound'' scale $E_{\text{QG}}$ of order
\begin{equation}
E_{\text{QG}} \sim M_4 / \sqrt{N(E_{\text{QG}} )}\,,~~~N(E_{\text{QG}} ) \sim E_{\text{QG}}  M_4 /\Lambda_4^{1/2}\,.
\end{equation}
This in turn implies
\begin{equation}
E_{\text{QG}}  \sim\Lambda_4^{1/6} M_4^{1/3} \sim 0.02 \text{ GeV}\,,
\end{equation}
which conflicts with the experimental fact that gravity remains weakly coupled at energies accessible at the LHC.

Our above analysis takes place entirely at a classical level. As shown in \eqref{paramred}, the cosmological constant $\Lambda_d$ acquires an exponential dependence on the radion $\lambda$. If the radion is not stabilized by quantum effects, the dimensionally reduced theory does not have a vacuum. Suppose that we now insist that the theory should in fact have a vacuum, so the radion must be stabilized.
This can be done most simply by including the 1-loop Casimir energy contributions to the radion potential from the light particles. Upon dimensional reduction to $d$ dimensions, the contribution from a particle of mass $m$ takes a very simple form in the massless limit ($m R \ll 1$) \cite{ArkaniHamed:2007gg}:
\begin{equation}
V_{C}(\lambda) = \mp \frac{2}{(2 \pi R)^{d} \Omega_{d}} \zeta(d+1) \e^{\frac{d(d-1)}{2(d-2)} \lambda} \,,~~~\Omega_d = \frac{2 \pi^{(d+1)/2}}{\Gamma(\frac{d+1}{2})}\,.
\label{VC}
\end{equation}
Here, $\zeta(x)$ is the Riemann zeta function, $\Omega_{d}$ is the volume of the unit $d$-sphere, and the $+$ sign is for bosons or fermions with antiperiodic boundary conditions, while the $-$ sign is for fermions with periodic boundary conditions. The 1-loop Casimir energy for a particle much heavier than $1/R$ is exponentially suppressed as $\e^{-2 \pi m R}$, so for our purposes, it suffices to consider contributions only from light particles with $m \lesssim 1/R$. For simplicity, we approximate all such particles to be massless.

The full potential at 1-loop, including both the classical contribution and the 1-loop Casimir energy from light particles, is then given by
\be
V(\lambda)  = V_\Lambda(\lambda) + V_C(\lambda) = (2 \pi R) \Lambda_D \e^{\frac{\lambda}{d-2}} - \sum_{n| m_n < 1/R} (-)^{F_n} \frac{2}{(2 \pi R)^{d} \Omega_{d}} \zeta(d+1)  \e^{\frac{d(d-1)}{2(d-2)} \lambda}\,.
\ee
We differentiate with respect to $\lambda$ and set this to 0 to find a critical point:
\be
0 = \partial_\lambda V(\lambda)  = \frac{1}{d-2} (2 \pi R) \Lambda_D \e^{\frac{\lambda}{d-2}} - \frac{d(d-1)}{2(d-2)} \sum_{n| m_n < 1/R} (-)^{F_n} \frac{2}{(2 \pi R)^{d} \Omega_{d}} \zeta(d+1)  \e^{\frac{d(d-1)}{2(d-2)} \lambda}\,.
\label{eq:diff}
\ee
The exponent of the second term is larger than that of the first term. This means that in order to find a minimum, the second term must be positive, while the first is negative. For $\Lambda_D < 0$, we may therefore obtain a minimum in $d$ dimensions provided that fermions with periodic boundary conditions dominate the Casimir energy. For $\Lambda_D > 0$, on the other hand, we may find a maximum in $d$ dimensions provided that bosons or fermions with antiperiodic boundary conditions dominate the Casimir energy. Creating a de Sitter minimum in this way would require a delicate interplay between bosons and fermions, and it would require us to go beyond the massless particle approximation we have employed here.

Let us now suppose that we have a tower of particles of equal spin $F$, with masses 
\be
m_n = n m \,,~~~~ n \in \mathbb{Z}_{>0}\,,
\ee
so that the sum in \eqref{eq:diff} becomes
\be
 \sum_{n| m_n < 1/R} (-)^{F_n} =  \sum_{n=1}^{1/(m R)} (-)^F = (-)^F \frac{1}{m R}\,.
 \label{eqLsum}
\ee
In order for the two terms in \eqref{eq:diff} to balance, we must therefore have (ignoring $O(1)$ factors):
\be
\frac{1}{mR^{D}} \sim |\Lambda_d|\,.
\ee
We further impose that the bound \eqref{massvcc} must be saturated both before and after reduction:
\be
m \sim |\Lambda_D|^{\alpha_D} M_D^{1 - 2 \alpha_D}  \sim |\Lambda_d|^{\alpha_d} M_d^{1 - 2 \alpha_d} \,.
\ee
Putting these together, we may eliminate $R$ and $m$ to find a relation between $\Lambda_D$ and $\Lambda_d$, which gives the following relation between $\alpha_D$ and $\alpha_d$:
\be
\alpha_D =  \alpha_d - \frac{2 \alpha_d^2 + \alpha_d - 1}{2 \alpha_d - d^2+ 3} \,.
\label{recursion}
\ee
This is a recursive relation between $\alpha_d$ and $\alpha_{d+1}$, which has a 1-parameter family of solutions. One such solution is the value $\alpha_d = 1/2$, which we found above from a strictly classical analysis. Another solution is $\alpha_d = 1/d$: this value is special in that it is the smallest possible value of $\alpha_d$ consistent with the recursion relation \eqref{recursion} that remains non-negative in the limit $d \rightarrow \infty$. Note that it is also the unique value of $\alpha_d$ for which the Planck scale $M_d$ drops out of \eqref{massvcc}, and in the reduction studied above it leads to $m \sim 1/R \sim |\Lambda_d|^{1/d} \sim |\Lambda_D|^{1/D}$, so the sum over $n$ in \eqref{eqLsum} runs over an $O(1)$ number of terms.

The bound $m \lesssim |\Lambda_d|^{1/d}$ is rather tantalizing, as the
current upper bound on the sum of the masses of the three standard model neutrinos is $0.3$ eV, coming from a combination of CMB measurements, galaxy surveys, and Lyman-alpha forest data. The lightest neutrino may in principle be massless, though there are lower bounds on the differences in the squared-masses of the neutrinos from observations of neutrino oscillations. This requires that at least one neutrino must have a mass above about $0.05$ eV.
The cosmological constant in our universe is measured to be
\begin{equation}
\Lambda_4^{1/4} \simeq 10^{-30} M_4 \simeq 0.002 \text{ eV}\,,
\end{equation}
which is quite close to the neutrino mass scale of roughly $0.01-0.1$ eV. As discussed in Section \ref{sec:INTRO}, the relationship between the neutrino mass and the cosmological constant implies a relationship between the Higgs vev and the cosmological constant, if we assume that the neutrino mass comes from the see-saw mechanism, thereby relating the cosmological constant problem to the hierarchy problem. This argument is reminiscent of the work of \cite{Ooguri:2016pdq, Ibanez:2017oqr,Ibanez:2017kvh}.

This application to our own universe comes with an important caveat, however: our 1-loop calculation produced minima of the radion potential only in the case of a negative cosmological constant $\Lambda_d < 0$, whereas our universe has $\Lambda_d > 0$, for which our simple analysis yields maxima of the potential. This could be remedied by moving beyond the massless limit and considering both fermions and bosons (or fermions with antiperiodic boundary conditions). Indeed, balancing the 1-loop Casimir energies of particles in the standard model leads to a vacuum in three dimensions upon dimensional reduction \cite{ArkaniHamed:2007gg}.


\section{Interlude: Eternal Inflation in Higher Dimensions}\label{sec:ETERNAL}

We now take a short break from our dimensional reduction analysis to work out the conditions for eternal inflation in general spacetime dimension $d$ by solving the Fokker-Planck equation, generalizing the computations of \cite{Rudelius:2019cfh} (see also \cite{Starobinsky:1986fx, Rey:1986zk, Linde:1991sk, Creminelli:2008es, Barenboim:2016mmw, Kinney:2018kew, Brahma:2019iyy, Seo:2020ger, Bedroya:2020rac, Blumenhagen:2020doa, Chojnacki:2021fag}).\footnote{We are very thankful to Liam McAllister for discussions on the computations in this section.} In Section \ref{sec:DS} below, we will comment briefly on the behavior of the conditions for avoiding eternal inflation under dimensional reduction.

We begin from the $d$-dimensional metric
\begin{equation}
ds^2 = -dt^2 + a(t)^2 d\vec{x}^2.
\end{equation}
and the action
\begin{equation}
S= \int d^d x \sqrt{-g} \left[ \frac{M_d^{d-2}}{2}  \mathcal{R} - \frac{1}{2} (\nabla \phi)^2 -  V(\phi)  \right].
\end{equation}
For simplicity, we take $\phi$ to be canonically normalized, and at 0th order we take it to be homogenous so that its spatial derivatives vanish. The equation of motion for the scalar field is thus
\begin{equation}
\ddot \phi + (d-1) H \dot \phi = - V'(\phi)\,,
\end{equation}
where $H = \dot a/a$.
The stress-energy tensor is given by
\begin{align}
T^\mu_{\nu} &= - g^{\mu \delta} \frac{2}{\sqrt{-g}} \frac{\delta S}{\delta g^{\delta\nu}}  = \text{diag}\left(-\frac{1}{2} \dot\phi^2-V(\phi), \frac{1}{2} \dot\phi^2-V(\phi),..., \frac{1}{2} \dot\phi^2-V(\phi)\right).
\end{align}
This is the stress-energy tensor of a perfect fluid of density $\rho$, pressure $p$, with
\begin{equation}
\rho = \frac{1}{2} \dot \phi^2 + V(\phi)\,,~~~ p  = \frac{1}{2} \dot \phi^2 - V(\phi)\,.
\end{equation}
Einstein's equations are given by
\begin{equation}
G_{\mu\nu}= \kappa_d^2 T_{\mu\nu} =  \frac{1}{M_d^{d-2}} T_{\mu\nu}\,.
\end{equation}
Here, we have
\begin{gather}
G_{00} = \frac{1}{2} (d-1)(d-2) H^2 = \frac{\rho}{M_d^{d-2}} \,,  \\
\frac{1}{a^2} G_{11} =...=\frac{1}{a^2} G_{d-1,d-1} = - (d-2)  {\ddot a \over a} - \frac{1}{2}(d-2)(d-3) H^2 = \frac{p}{M_d^{d-2}}\,. 
\end{gather}
This gives the Friedmann equations:
\begin{equation}
H^2 = \frac{2 \rho}{(d-1)(d-2) M_d^{d-2}} \,,~~~\frac{\ddot a}{a} = - \frac{p}{(d-2) M_d^{d-2}} - \frac{(d-3) H^2}{2 }\, .
\end{equation}
Inflation takes place in the slow-roll regime, 
\be
\dot \phi^2 \ll V(\phi) \,,~~~~~|\ddot \phi | \ll H |\dot \phi| ,|V'(\phi)| \,,
\ee
in which case the equation of motion for $\phi$ and the first Friedmann equation become
\be
(d-1) H \dot \phi = - V'(\phi) \,,~~~~ H^2 = \frac{2 V(\phi)}{(d-1)(d-2) M_d^{d-2}} \,.
\label{slowroll}
\ee

To study eternal inflation, we further need to incorporate backreaction from quantum fluctuations of the scalar field. The scalar 2-point function in dS$_d$ takes the form \cite{Dubovsky:2011uy}
\be
\langle \phi^2 \rangle = \frac{H^{d-1}}{ \pi \Omega_{d-2}} t + \ldots \,,~~~~~~\Omega_{d-2} = \frac{2 \pi^{(d-1)/2}}{\Gamma(\frac{d-1}{2})}\,.
\label{eq:2pt}
\ee
where $\ldots$ represents terms that are not linear in $t$. This linear term encodes the effect of Gaussian quantum fluctuations that exit the horizon, decohere, and backreact on the classical slow-roll equation of motion \eqref{slowroll}. This backreaction takes the form 
\be
(d-1) H \dot \phi  + V'(\phi)  = N(t)\,,
\ee
where $N(t)$ is a Gaussian noise term, which induces a random walk of the field $\phi$ in the potential $V(\phi)$. In other words, in an infinitesimal time $\delta t$, $\phi$ will vary according to
\be
\delta \phi = -\frac{1}{(d-1) H } V'(\phi) \delta t  + \delta \phi_q(\delta t)\,,~~~~~\delta \phi_q(\delta t) \sim \mathcal{N}(0, \frac{H^{d-1}}{ \pi \Omega_{d-2}} \delta t)\,,
\ee
where the variance of the normal distribution comes from the coefficient of the linear term in \eqref{eq:2pt}.

If we further approximate $H$ as a constant, which is well-justified in the slow-roll regime, the evolution of the probability distribution $P[\phi, t]$ of $\phi$ as a function of $t$ is then described by a Fokker-Planck equation \cite{Starobinsky:1986fx, Rey:1986zk, Linde:1991sk}:
\be
\dot P[\phi, t] = \frac{1}{2} \left(\frac{H^{d-1}}{ \pi \Omega_{d-2}} \right) \partial_\phi^2 P[\phi, t] + \frac{1}{(d-1) H} \partial_\phi \Big( (\partial_\phi V(\phi)) P[\phi,t] \Big)\,.
\ee
This equation is difficult to solve for a general potential $V(\phi)$, but the solution takes a simple, analytic form when the potential is linear or quadratic in $\phi$. In both cases, the solution takes the form of a Gaussian distribution:
\be
P[\phi,t] = \frac{1}{ \sigma(t) \sqrt{2 \pi}} \exp\left[ - \frac{ (\phi - \mu(t))^2}{ 2 \sigma^2(t) } \right]\,.
\label{eq:Gaussian}
\ee
For a linear potential $V = V_0 - \alpha \phi$, the parameters $\mu(t)$, $\sigma^2(t)$ are given by
\be
\mu(t) = \frac{\alpha}{(d-1) H} \, t\,,~~~~\sigma^2(t) = \frac{H^{d-1}}{ \pi \Omega_{d-2}}\,  t\,.
\label{eq:linearevolution}
\ee
For a quadratic hilltop potential $V = V_0 -  \frac{1}{2} m^2 \phi^2$, the parameters are instead given by
\be
\mu(t) = 0 \,,~~~~ \sigma^2(t) = \frac{(d-1) H^d}{2 \pi \Omega_{d-2} m^2 }\left[-1 + \exp\left(\frac{2 m^2  }{(d-1) H} \, t \right) \right] \,.
\label{tachyonevolution}
\ee
For the linear potential, inflation occurs if $\phi < \phi_c$, where $\phi_c$ is some critical value whose precise value will be unimportant to us. The probability that $\phi  < \phi_c$ at time $t$ is given by
\begin{equation}
\text{Pr}[\phi> \phi_c, t] = \int_{-\infty}^{\phi_c} d\phi \,P[\phi,t],
\end{equation}
where $P[\phi,t]$ is the probability density function for a Gaussian distribution, given in (\ref{eq:Gaussian}), with mean $\mu(t)$ and variance $\sigma^2(t)$ given by (\ref{eq:linearevolution}). The result is
\begin{equation}
\text{Pr}[\phi> \phi_c, t]  = \frac{1}{2} \text{erfc}\left[\frac{\mu(t)-\phi_c}{\sigma(t) \sqrt{2} } \right] =  \frac{1}{2} \text{erfc}\left[ \frac{\frac{\alpha }{(d-1) H}t-\phi_c}{ \left( \frac{ 2H^{d-1}t }{ \pi \Omega_{d-2}} \right)^{1/2} } \right],
\end{equation}
with erfc the error function. For large $t$, this error function can be approximated to leading order as an exponential,
\begin{equation}
\text{Pr}[\phi> \phi_c, t] \sim \exp \left[ - \left( \frac{\frac{\alpha }{(d-1) H}t-\phi_c}{ \left( \frac{ 2H^{d-1}t }{ \pi \Omega_{d-2}} \right)^{1/2} }  \right)^2 \,\right] \sim \exp \left[ -   \frac{\pi \Omega_{d-2} \alpha^2}{2 (d-1)^2 H^{d+1}} t \right],
\label{eq:lasteq}
\end{equation}
This means that the probability of inflation occurring at time $t$ for a fixed comoving observer decays exponentially with time. On the other hand, the volume of the inflating region grows exponential in time as $\exp((d-1) H t)$. Eternal inflation occurs if this exponential growth beats the exponential decay, i.e., if
\begin{equation} 
(d-1) H  > \frac{\pi \Omega_{d-2} \alpha^2}{2 (d-1)^2 H^{d+1}}\,.
\end{equation} 
Substituting $V$ for $H$ using \eqref{slowroll} and setting $\alpha =  \nabla V(\phi)$, this becomes
\begin{equation}
\frac{|\nabla V|^2}{V^{(d+2)/ 2} } <  \frac{2 (d-1)^3}{\pi \Omega_{d-2} } \left(  \frac{ 2  }{ (d-1)(d-2) M_d^{d-2} }   \right)^{(d+2)/2}\,.
\label{eq:lineareternal}
\end{equation}

A similar analysis applies to the quadratic hilltop potential. Now, inflation occurs if $|\phi| < \phi_c$, and the probability of this occurring at time $t$ is given by
\begin{equation}
\text{Pr}[|\phi| < \phi_c, t] = \int_{-\phi_c}^{\phi_c} d\phi \,P[\phi,t]\,,
\end{equation}
The probability density function is a Gaussian with mean and variance given by (\ref{tachyonevolution}). This gives
\begin{align}
\text{Pr}[|\phi| < \phi_c, t]   = \text{erf}\left[\frac{\phi_c-\mu(t)}{\sigma(t) \sqrt{2}} \right] &=  \text{erf}   \left[  \left( \frac{   \pi \Omega_{d-2}   m^2  \phi_c^2 }{ (d-1) H^d  \left(-1 + \exp\left(\frac{2 m^2  }{(d-1) H} \, t \right) \right) } \right)^{1/2}    \right]  \nonumber \\
&  \sim  \exp\left[ - \frac{m^2  }{(d-1) H}  \, t\right]\,,
\label{decay}
\end{align}
where in the last line we have Taylor-expanded the error function at the origin.
As before, eternal inflation occurs if the exponential expansion of the universe dominates the exponential decay of \eqref{decay}, which is equivalent to the condition
\begin{equation}
(d-1) H > \frac{m^2}{(d-1) H}\,.
\end{equation}
Substituting $V$ for $H$ using \eqref{slowroll} and setting $m^2 = -V''$, this becomes
\begin{equation}
\frac{V''}{V} > -\frac{2 (d-1)}{d-2} \frac{1}{M_d^{d-2}} \,.
\label{eq:hilltopeternal}
\end{equation}

As in \cite{Rudelius:2019cfh}, this analysis may be generalized straightforwardly to theories with multiple scalar fields: as long as the potential $V(\phi^i)$ separates into a sum $\sum_i V_i(\phi^i)$, where each $V_i$ is linear or quadratic and depends only on $\phi^i$, the solution to the Fokker-Planck equation will separate into a product of Gaussian wavepackets. The main upshot of this is that, at a critical point of a multifield potential, one should replace $V''$ in \eqref{eq:hilltopeternal} with a sum over the negative eigenvalues of the Hessian. 

To conclude this section, let us compare the bound \eqref{eq:hilltopeternal} to the analogous bound in the Refined de Sitter Conjecture (RdSC) \cite{Garg:2018reu, Ooguri:2018wrx, Andriot:2018mav}:
\begin{equation}
\frac{\text{min} (\nabla_i \nabla_j V)}{V} \leq -\frac{c_d' }{M_d^{d-2}}\,,
\label{eq:RdSC}
\end{equation}
where $c_d'$ is some $O(1)$ constant, and $\text{min} (\nabla_i \nabla_j V)$ is the minimum eigenvalue of the Hessian. We see here that the RdSC is incompatible with eternal inflation provided $c_d' > 2 (d-1)/(d-2) $. The RdSC bound with this value of $c_d'$ seems to be violated in some 4d examples with 10d supergravity uplifts \cite{Andriot:2020wpp, Andriot:2021rdy}, but these solutions do not lie in the classical regime of string theory, so they may be modified by stringy effects \cite{Andriot:2020vlg}. Furthermore, even if the RdSC bound with this value of $c_d'$ is violated for the smallest eigenvalue of the Hessian, one might still violate the necessary conditions for eternal inflation upon summing over the other negative eigenvalues of the Hessian.


\section{Derivatives of Scalar Field Potentials}\label{sec:DS}

Next, we consider a bound on scalar field potentials of the form
\be
\frac{ |\nabla V_D|  }{ V_D^{\gamma_D}  }  \geq c_D M_D^{D(1-\gamma_D)-(D-2)/2}\,,
\label{eq:DdSC}
\ee
for $V_D > 0$, where $\gamma_D$ and $c_D$ are $O(1)$ constants.
Here, $|\nabla V_D|^2 = G^{ij} \partial_i V_D \partial_j V_D$, and the action takes the form
\begin{align}
S =  \int \rmd^D x \sqrt{-g}  \left[ \frac{M_D^{D-2}}{2} {\cal R}_D -  \frac{1}{2} G_{ij}(\phi) \nabla \phi_i \nabla \phi_j   - V_D(\phi) \right] \,.
 \label{eq:action}
\end{align}
We want to understand how this behaves under dimensional reduction. We take our usual dimensional reduction ansatz:
\begin{equation}
ds^2 = e^{\lambda/(d-2)} d \hat{s}^2 + e^{-\lambda} dy^2\,.
\end{equation}
The resulting action in $d$ dimensions is given by 
\be
S = \int \rmd^d x \sqrt{- \hat g} \left[  \frac{M_d^{d-2}}{2 } \left( {\cal R}_d  -  \frac{d-1}{4(d-2)} (\nabla \lambda)^2 \right)  -  \frac{1}{2} G_{ij}^{(d)}(\phi) \nabla \phi_i \nabla \phi_j  -  V_d(\phi) \right] \,,
 \label{eq:dSactiond}
\ee
where we have
\begin{align}
M_d^{d-2} = (2 \pi R) M_D^{D-2}\,,~~~G_{ij}^{(d)} = (2 \pi R) G_{ij}^{(D)} \,,~~~
V_d = V_D (2 \pi R) e^{\lambda/(d-2)} \, .
\end{align}
There is also a contribution to $V_d$ from the Casimir energy, but for large $R$, this will be parametrically subdominant to the classical term, and we will neglect it here.

Thus we have
\begin{align}
|\nabla V_d|^2 &= (G^{(d)})^{ij} \partial_i V_d \partial_j V_d + \frac{4(d-2)}{d-1} \frac{1}{M_d^{d-2}} (\partial_\lambda V_d)^2 \, , \nonumber \\
&= (2 \pi R) (G^{(D)})^{ij} \partial_i V_D \partial_j V_D + \frac{4}{(d-1)(d-2)} \frac{2 \pi R}{M_D^{D-2}} V_D^2 \,, \label{nablaVred} \\
&= (2 \pi R) |\nabla V_D|^2 + \frac{4}{(d-1)(d-2)} \frac{2 \pi R}{M_D^{D-2}} V_D^2 \nonumber \,,
\end{align}
where we have set $\langle \lambda \rangle = 0$ after taking the $\lambda$ derivative. This yields
\begin{equation}
\frac{|\nabla V_d|^2}{V_d^{2 \gamma_d}} M_d^{d-2- 2 d(1-\gamma_d)} = (2 \pi R)^{\frac{4 }{d-2}(\gamma_d-1)} M_D^{\frac{d-1}{d-2}(2 d \gamma_d -d-2)} \frac{|\nabla V_D |^2 + \frac{4}{(d-1)(d-2)} M_D^{2-D} V_D^2 }{V_D^{2 \gamma_d}}\,.
\end{equation}
From this, we see that the value $\gamma_D = \gamma_d = 1$ is distinguished by dimensional reduction: for this particular value, the $R$-dependence cancels, and if we ignore the $\lambda$ dependence of $V_d$, the bound \eqref{eq:DdSC} is exactly preserved under dimensional reduction. With $\gamma_D = 1$, this bound matches the ``de Sitter Conjecture'' bound of \cite{Obied:2018sgi}.

Setting $\gamma_D = \gamma_d = 1$, we may further fix the constant $c_D$ by including the $\lambda$ dependence of $V_d$. In particular, assuming that the bound \eqref{eq:DdSC} is exactly preserved under dimensional reduction, so that $V_D$ saturates the bound in $d$ dimensions precisely when $V_d$ saturates the bound in $d$ dimensions, we have
\begin{equation}
c_d^2 = \frac{|\nabla V_d|^2}{V_d^2} M_d^{d-2} =  \frac{ c_D^2 V_D^2 + \frac{4}{(d-1)(d-2)} V_D^2 }{V_D^2}  \, ,
\end{equation}
where in the first equality we have assumed that the bound is saturated $d$ dimensions, and in the second we have used \eqref{nablaVred} and assumed that it is saturated in $D$ dimensions. This gives a recursive relation for $c_D$:
\begin{equation}
c_d^2 =  c_D^2 +  \frac{4}{(d-1)(d-2)}\,,
\end{equation}
which is solved by
\begin{equation}
c_d^2 = \beta + \frac{4}{d-2}\,,
\end{equation}
where $\beta$ is a free parameter. This leads to the bound
\be
\frac{|\nabla V_d|}{V_d} \geq  \sqrt{\beta + \frac{4}{d-2}}  \frac{1}{M_d^{(d-2)/2}}\,.
\label{ourbounda}
\ee
Our dimensional reduction argument leaves $\beta$ unfixed, but the value $\beta = 0$ is distinguished for two reasons: first of all, it is the smallest value of $\beta$ such that $c_d$ remains non-negative in the limit $d \rightarrow \infty$. Secondly, the bound with $\beta = 0$,
\be
\frac{|\nabla V_d|}{V_d} \geq  \sqrt{\frac{4}{d-2}}  \frac{1}{M_d^{(d-2)/2}}\,,
\label{ourbound}
\ee
 has physical meaning: it is exactly the condition required for a theory of Einstein gravity coupled to a dilaton field to avoid accelerated expansion of the universe in $d$ dimensions \cite{Obied:2018sgi}.
 
Observational constraints rule out single-field quintessence models satisfying \eqref{ourbound} \cite{Agrawal:2018own}. Even if one entertains the possibility of fine-tuned, multi-field quintessence models \cite{Cicoli:2020noz, Akrami:2020zfz, Farakos:2021mwt}, it is very hard to imagine that \eqref{ourbound} could be satisfied at the maximum of the Higgs potential \cite{Denef:2018etk}. However, this bound may yet be satisfied in asymptotic limits of scalar field space, as is evidenced by many examples in string theory. For instance, toroidal compactifications of $O(16) \times O(16)$ heterotic string theory to $d$ dimensions satisfy the bound \cite{Obied:2018sgi}:
\be
\frac{|\nabla V_d|}{V_d} \geq \text{min}\left(2 \sqrt{ \frac{3d -5}{ d-2 }  }, \frac{4 \sqrt{2}}{  \sqrt{ (10-d) (d-2)  } }\right)  \frac{1}{M_d^{(d-2)/2}}  \,.
\ee
This bound implies \eqref{ourbound}, and it implies \eqref{ourbounda} in $d \geq 3$ provided $\beta \leq 4/7$.

Next, we may consider the KKLT scenario \cite{Kachru:2003aw} and the LVS scenario \cite{Conlon:2005ki}.  Both of these models have de Sitter critical points, so they violate \eqref{ourbound} in the interior of scalar field space. However, as noted in \cite{Bedroya:2019snp}, the KKLT potential behaves asymptotically as
\be
V(\phi) \sim \exp \left( - \sqrt{6} \frac{ \phi }{ M_4} \right)\,,
\ee
and the LVS potential behaves asymptotically as
\be
V(\phi) \sim \exp \left( -3 \sqrt{ \frac{3}{2} } \frac{\phi }{ M_4 }\right)\,.
\ee
Thus, in asymptotic limits of field space, both of these satisfy \eqref{ourbound}, and more generally the former satisfies \eqref{ourbounda} provided $\beta \leq 4$, whereas the latter requires $\beta \leq 23/2$. More examples of 4d theories satisfying \eqref{ourbound} in asymptotic regions of scalar field space can be found in \cite{Obied:2018sgi, Andriot:2020lea,  Andriot:2021rdy, Wrase:2010ew, Andriot:2019wrs}. 

However, some of the string compactifications considered in \cite{Wrase:2010ew, Andriot:2019wrs, Andriot:2020lea} naively seem to violate \eqref{ourbound}. For instance, in Type IIA compactifications in the presence of O4-planes but no D4-brane sources, \cite{Wrase:2010ew} derived a lower bound on $|\nabla V|  M_4 /V$ of $\sqrt{2/3}$, which if saturated would violate \eqref{ourbound} while saturating the Transplackian Censorship Conjecture (TCC) bound \cite{Bedroya:2019snp}:
\be
\frac{|V'|}{V} \geq \frac{2}{\sqrt{(d-1)(d-2)}}  \frac{1}{M_d^{(d-2)/2}}\,.
\label{TCC}
\ee
However, there is an important subtlety here: the lower bound in this example comes from considering only the volume modulus and the dilaton, so it is plausible that including other scalar fields could lead to consistency with \eqref{ourbound} in asymptotic limits of scalar field space. Likewise, the lower bounds on  $|V'|  M_4 /V$ listed in Table 2 of \cite{Andriot:2020lea} come from considering the derivative $V'$ of the potential with respect to the geodesic distance along certain geodesics in scalar field space, not from considering the gradient $\nabla V$ of the potential with respect to \emph{all} scalar fields in the theory.\footnote{The fact that the derivative with respect to geodesic distance $V'$ may saturate the TCC bound \eqref{TCC} is crucial for the connection between the TCC and the Swampland Distance Conjecture (SDC) proposed in \cite{Andriot:2020lea}.} Once contributions to the gradient $\nabla V$ from additional scalar fields are included, it is plausible that our bound \eqref{ourbound} may yet be satisfied.\footnotemark

\footnotetext{We are very thankful to David Andriot for discussions on these points.}

It may sound like wishful thinking to hope that contributions to the gradient from additional scalar fields will always come to the rescue, ensuring consistency with \eqref{ourbound}.
However, we already know of at least one simple example where this is precisely what happens: in heterotic string theory compactified to four dimensions, one has $ |\partial_\rho V| M_4/V = \sqrt{2/3}$, where $\rho$ is the volume modulus of the compactification manifold (see Table 2 of \cite{Andriot:2020lea}). This saturates the TCC bound \eqref{TCC} along a geodesic in the $\rho$ direction and naively violates our bound \eqref{ourbound}. However, the potential also depends asymptotically on the dilaton, $ |\partial_\tau V| M_4 /V = \sqrt{2}$. This additional contribution to the gradient leads to consistency with \eqref{ourbound}.

Note, however, that while this example is a useful proof of principle, it does not represent the typical scenario. In general, it is quite difficult to determine the dependence of the potential on \emph{every} scalar field in the theory, so falsifying \eqref{ourbound} in asymptotic regions of scalar field space is not easy to do.\footnotemark[\value{footnote}]

Finally, note that \eqref{ourbound} is incompatible with the condition \eqref{eq:lineareternal} needed for eternal inflation, since $\gamma_d = 1 < \gamma_d^{\text{EI}} = (d+2)/4$. This is rather unsurprising from a physical perspective: \eqref{ourbound} forbids accelerated expansion of the universe, which is obviously a prerequisite for eternal inflation. Dimensional reduction induces an exponential potential for the radion $\lambda$, $V_d = V_D \exp(\lambda/(d-2))$, so if the radion is not stabilized, this exponential potential alone will suffice to violate the condition \eqref{eq:lineareternal} required for eternal inflation. Radion stabilization will not occur in asymptotic regions of scalar field space \cite{Dine:1985he}, so dimensional reduction will not lead to eternal inflation in such regions.


\section{Conclusions}\label{sec:CONC}

We have seen that dimensional reduction distinguishes particular $O(1)$ constants appearing in the WGC, RFC, (A)dSDC, and dSC. In the case of the (A)dSDC, one mass scale that emerges is suggestively close to the neutrino mass scale, which may be related to the Higgs vev via the see-saw mechanism. It might be worthwhile to search for other swampland-related reasons for a connection between the neutrino mass/Higgs vev and the cosmological constant.

In the case of the dSC, the bound we obtain precisely forbids accelerated expansion of the universe in Einstein-dilaton gravity. This is interesting in that it offers an alternative physical explanation to Trans-Planckian Censorship \cite{Bedroya:2019snp} as to why the dSC seems to hold universally in asymptotic limits of scalar field space: perhaps quantum gravity simply forbids accelerated expansion in such regions. This possibility merits further investigation.

We have also worked out conditions for eternal inflation in $d \geq 4$ spacetime dimensions, generalizing previous bounds obtained in $d=4$. Models of inflation in more than four dimensions have not received very much attention, due to the obvious fact that such models are of little experimental interest. But given the recent, renewed interest in scalar field potentials and de Sitter vacua in quantum gravity, it may be worthwhile to temporarily ignore the question of experimental relevance and instead explore more general theories of inflation and cosmology in diverse dimensions, as well as their possible embeddings in string theory. It would be interesting to study de Sitter critical points of scalar field potentials in string compactifications to $d > 4$ dimensions, to see if these critical points may support eternal inflation.

We have not answered the most important question: to what extent are the bounds we have obtained actually satisfied in quantum gravities? It is clear that many string compactifications satisfy them, but it is also quite possible that these results suffer from a lamppost effect. Understanding the domain of validity of these conjectured bounds is a crucial task for the swampland program, though it may not be an easy one.


\section*{Acknowledgements}

We thank David Andriot, Ben Heidenreich, Juan Maldacena, Liam McAllister, Georges Obied, Eran Palti, Matthew Reece, and Cumrun Vafa for useful discussions. We thank David Andriot, Ben Heidenreich, Liam McAllister, and Matthew Reece for comments on a draft of this paper. We acknowledge hospitality from the 2019 Simons Summer Workshop at the Simons Center for Geometry and
Physics at Stony Brook University, where some of these discussions took place. The work of TR was supported by NSF grant  PHY1820912, the Simons Foundation, and the Berkeley Center for Theoretical Physics.

\bibliographystyle{utphys}
\bibliography{ref}

\end{document}